\documentclass[aps,pre,preprint,nofootinbib,eqsecnum,tightenlines,
showpacs,amsmath,titlepage]{revtex4}
\usepackage{graphicx}
\begin{document}

\vspace*{2cm}

\title{Towards a unification of hierarchical reference theory and
       self-consistent Ornstein-Zernike approximation: Analysis of
       exactly solvable mean-spherical and generalized mean-spherical
       models}

\author{J. S. H{\o}ye}\email{johan.hoye@phys.ntnu.no}
\author{A. Reiner}\email{areiner@tph.tuwien.ac.at}

\affiliation{Department of Physics, Norwegian University of Science and
Technology, N-7491 Trondheim, Norway}

\date{\today}

\begin{abstract}

The hierarchical reference theory (HRT) and the self-consistent
Ornstein-Zernike approximation (SCOZA) are two liquid state theories
that both furnish a largely satisfactory description of the critical
region as well as the phase coexistence and equation of state in
general. Furthermore, there are a number of similarities that suggest
the possibility of a unification of both theories. Earlier in this
respect we have studied consistency between the internal energy and
free energy routes. As a next step toward this goal we here consider
consistency with the compressibility route too, but we restrict
explicit evaluations to a model whose exact solution is known showing
that a unification works in that case. The model in question is the
mean spherical model (MSM) which we here extend to a generalized MSM.
For this case, we show that the correct solutions can be
recovered from suitable boundary conditions through either of SCOZA or
HRT alone as well as by the combined theory.  Furthermore, the
relation between the HRT-SCOZA equations and those of SCOZA and HRT
becomes transparent.

\end{abstract}


\maketitle

\section{Introduction}
\label{intro}

Both the self-consistent Ornstein-Zernike approximation (SCOZA)
\cite{hoyestell84, hoyestell85, hoyestell77} and the hier{\-}archical
reference theory (HRT) \cite{parolareatto95} have been found to
give very accurate results for fluids in thermal equilibrium. In
particular, the respective non-linear partial differential equations
can be solved in the critical region, and their solution gives
non-classical, and partly Ising-like, critical indices. These
equations are derived by obtaining the equation of state in two
independent ways and using thermodynamic consistency to fix a free
parameter in the direct correlation function.

Although both approaches appear similar in various ways, there are
also marked differences. Both approaches make use of the
compressibility route to thermodynamics, but SCOZA combines it with
the internal energy route while HRT, inspired by momentum-space
renormalization group theory, uses the Helmholtz free energy
route. Thus, in short, the SCOZA adds effective strength to the
attractive interaction by increasing inverse temperature $\beta=1/k_B
T$ while HRT adds contributions to the interaction by including its
Fourier components for shorter wavenumbers $Q$ until the limit of
interest $Q\rightarrow 0$ is obtained.

In a recent work we considered thermodynamic consistency between the
internal energy and free energy routes to thermodynamics
\cite{reinerhoye05}. In the present work we want to extend this to
consistency with the compressibility route, too. This requires the
introduction of two free parameters instead of a single one in each of
the original theories.  In view of the high accuracy of HRT and SCOZA,
one may expect this increased freedom to give even better results both
for spin systems and for fluids. Due to the complexity of the combined
problem, we here limit ourselves to a simpler situation that can be
analysed explicitly and for which the exact solution can be
established. This is the case for the mean spherical model (MSM). This
model can be considered as the limit $D\rightarrow\infty$ for
$D$-dimensional spins in $d$ space dimensions where the transverse
susceptibility is the relevant one for the fluctuation theorem or
compressibility relation. In this connection we realize that the MSM
can be extended in a straightforward way to a generalized MSM (GMSM)
that yields the same HRT and SCOZA problems as the MSM; the only
difference lies in the reference system boundary conditions. In the
former the spin length is fixed to~1 while in the latter the spin
length has some distribution of spin lengths.  It should be pointed
out that neither SCOZA nor HRT, nor our combined theory, are
restricted to simple fluids and their lattice gas version although
they are most often applied to these systems. The usual lattice gas is
equivalent to the Ising model with spins $s=\pm1$. So what we do here
is to generalize and apply both of these theories to continuous spins
on a lattice too.

In Sec.~\ref{sec2} we briefly consider the MSM, and in Sec.~\ref{sec3}
we extend it to a GMSM whose solution is established. In
Sec.~\ref{sec4} the SCOZA problem for the GMSM model is considered
while in Sec.~\ref{sec5} the corresponding HRT problem is
considered. With both of these approaches one free parameter can be
determined. By appropriate choice of this parameter we get partial
differential equations whose solutions are those of the GMSM with the
reference system as boundary condition. In Sec.~\ref{sec6} an
alternative method of solution based on a result from
Ref.~\onlinecite{reinerhoye05} is used.  Then unification of SCOZA and
HRT is considered in full generality in Sec.~\ref{sec7}. In
Sec.~\ref{sec8} a pair correlation function of MSM form containing two
free parameters is proposed and explicit HRT-SCOZA equations are
established for the GMSM situation where transverse susceptibility
replaces susceptibility. By analysis of these equations we show how
the HRT-SCOZA equations work in this case and how the GMSM solution is
recovered.

\section{Mean-spherical model}
\label{sec2}

Consider $D$-dimensional spins on a lattice in $d$ space dimensions
and with cells of unit volume.  It is well established that in the
limit $D\rightarrow\infty$ this statistical mechanical system can be
solved exactly \cite{stanley68, stanleykac68}. Its solution is the
same as that of the spherical model \cite{berlinkac52}. Here we will
consider its variant, the mean-spherical model (MSM)
\cite{lewiswannier52}. In the MSM the Ising spins are replaced by
interacting spins $s_i$ whose length is Gaussian distributed such that
$\langle s_i^2\rangle=1$. This is nothing but a Gaussian model with an
adjustable one-particle harmonic potential to keep $\langle
s_i^2\rangle=1$ fixed. More precisely one Laplace-transforms the
spherical constraint by which a Gaussian model partition function is
obtained.  In the thermodynamic limit (i.e., for an infinite system)
the inverse transform is determined by the maximum term of the
integrand. Following the evaluations by H{\o}ye and Stell the
resulting Gibbs free energy $g$ per spin becomes \cite{hoyestell97}
\begin{equation}
L=-\beta g=s+\frac{(\beta H)^2}{2(2s-\beta \tilde{\psi}(0))} +\frac{1}{2}\ln\pi-\frac{1}{2}\frac{1}{(2\pi)^d}\int\ln{(2s-\beta{\tilde\psi}(k))}\,d{\bf k}.
\label{1}
\end{equation}
Here $s$ is the Laplace-transform variable, $H$ is magnetic field, and
${\tilde\psi}(k)$ is the Fourier transform of the pair interaction
normalized to ${\tilde\psi}(0)=1$. The maximum of expression
(\ref{1}) is obtained by taking $\partial L/\partial s=0$. Then the
equation of state follows easily by utilizing the condition for
maximum. However, in the next section we will generalize the MSM so we
will come back to the equation of state there.

\section{Generalized mean-spherical model.}
\label{sec3}

In the MSM the spherical constraint $\langle s_i^2\rangle=1$ is
fixed. For $D$-dimensional spins ($D\rightarrow\infty$) this also
corresponds to spins of fixed length. Now we can generalize this and
let the $D$-dimensional spins have a distribution of lengths by which
the spherical constraint is removed. At thermal equilibrium the
average spin length squared will then change with both magnetization
and temperature. This variation will depend upon the equation of state
or the spin distribution specified for non-interacting spins. In the
limit $D\rightarrow\infty$ this model will again be exactly solvable
as the spin distribution becomes Gaussian, but the width of this
distribution varies both with temperature and magnetization.

With this in mind we first generalize the MSM to $\langle
s_i^2\rangle=n$ which replaces the first term of expression (\ref{1})
by $sn$. Further, as $n$ will not be fixed there must be a function
$F(n)$ that accounts for the distribution of $n$-values. In this way
expression (\ref{1}) can be generalized to
\begin{equation}
L=sn+\frac{(\beta H)^2}{2(2s-\beta \tilde{\psi}(0))} +\frac{1}{2}\ln\pi-\frac{1}{2}\frac{1}{(2\pi)^d}\int\ln{(2s-\beta{\tilde\psi}(k))}\,d{\bf k}-\frac{1}{2}F(n).
\label{2}
\end{equation}

Again in the termodynamic limit the free energy is determined by
maximum of expression (\ref{2}), but now with respect to both $s$ and
$n$. This gives the conditions
\begin{eqnarray}
\nonumber
\frac{\partial L}{\partial s}=n-\left(\frac{\beta H}{2s-\beta}\right)^2-\frac{1}{(2\pi)^d}\int\frac{d{\bf k}}{2s-\beta{\tilde\psi}(k)}=0\\
\frac{\partial L}{\partial n}=s-\frac{1}{2}F'(n)=0.\hspace*{3cm}
\label{3}
\end{eqnarray}
With this the magnetization becomes
\begin{equation}
m=\frac{\partial L}{\partial(\beta H)}=\frac{\beta H}{2s-\beta}.
\label{4}
\end{equation}
Now we  put
\begin{equation}
z=\frac{\beta}{2s}\quad{\rm and}\quad P(z)=\frac{1}{(2\pi)^d}\int\frac{d{\bf k}}{1-z{\tilde\psi}(k)},
\label{5}
\end{equation}
and Eqs.~(\ref{3}) and (\ref{4}) can be written
\begin{eqnarray}
\nonumber
\beta(n-m^2)=zP(z)\\
\frac{z}{\beta}=f(n)
\label{6}
\end{eqnarray}
with $f(n)=1/F'(n)$ and
\begin{equation}
\beta H=m(2s-\beta)=\frac{m}{n-m^2}P(z)-\beta m=m\left(\frac{\beta}{z}-\beta\right).
\label{7}
\end{equation}

{}From this we obtain the transverse susceptibility $\chi_\perp$ as
(for $D$-dimensional spins) \cite{hoyestell97}
\begin{equation}
\frac{\beta}{\chi_\perp}=\frac{\partial(\beta H_\perp)}{\partial m_\perp}=\frac{\beta H}{m}=\frac{\beta}{z}-\beta
\label{8}
\end{equation}
where $H_\perp\;(\rightarrow 0)$ and $m_\perp\;(\rightarrow 0)$ are
transverse magnetic field and transverse magnetization. The internal
energy (from pair interactions) becomes
\begin{eqnarray}
\nonumber
U=-\left(\frac{\partial L}{\partial \beta}\right)_H-mH=
-\frac{1}{2}\frac{1}{(2\pi)^d}\int\frac{{\tilde\psi}(k)\,d{\bf k}}{2s-\beta{\tilde\psi}(k)}-\frac{1}{2}\left(\frac{\beta H}{2s-\beta}\right)^2\\
=-\frac{1}{2}m^2-\frac{1}{2\beta}(P(z)-1).
\label{9}
\end{eqnarray}
In accordance with this the spin correlation function for transverse
correlations is
\begin{equation}
{\tilde\Gamma_\perp}(k)=\frac{z}{\beta(1-z{\tilde\psi(k)})}.
\label{10}
\end{equation}
Further in accordance with the fluctuation theorem 
${\tilde\Gamma_\perp}(0)=\chi_\perp/\beta$ which is consistent with
Eq.~(\ref{8}). By integrating this using (\ref{5}) and (\ref{6}) one
finds the ``core'' condition
\begin{equation}
\Gamma_\perp(0)=\frac{z}{\beta}P(z)=n-m^2.
\label{11}
\end{equation}
With $n=1$ fixed one is back to the constraint of the usual MSM.

For $n$ not fixed the function $f(n)$ must be defined by or related to
the reference system at $\beta=0$ where $P(z)=1$ as then $z\rightarrow
0$. So with Eqs.~(\ref{6})--(\ref{8}) and (\ref{11}) we have for the
reference system
\begin{equation}
\mu=\mu(m^2)=\frac{\chi_\perp}{\beta}=n-m^2=\frac{z}{\beta}=f(n).
\label{12}
\end{equation}
Thus
\begin{equation}
n=n(m^2)=\mu(m^2)+m^2.
\label{13}
\end{equation}
For general $\beta$ Eq.~(\ref{6}) then means
\begin{equation}
\mu=\mu(m^2)=n-m^2=\frac{z}{\beta}P(z)=f(n)P(z).
\label{14}
\end{equation}
Now we can put
\begin{equation}
\nonumber
P(z)=1+2J\quad{\rm and}\quad \mu_e=\frac{z}{\beta}=f(n)
\end{equation}
by which Eq.~(\ref{14}) becomes
\begin{eqnarray}
\nonumber
n-m^2=(1+2J)\mu_e=f(n)+2\mu_e J,\\
n-m_e^2=f(n)=\mu_e
\label{15}
\end{eqnarray}
where
\begin{equation}
m_e^2=m^2+2\mu_e J.
\label{16}
\end{equation}
Comparing Eqs.~(\ref{15}) and (\ref{16}) with Eqs.~(\ref{12}) and
(\ref{13}) one sees that this implies
\begin{equation}
\mu_e=\mu_e(m^2)=n(m_e^2)-m_e^2=\mu(m_e^2).
\label{17}
\end{equation}

\section{SCOZA for the model.}
\label{sec4} 

In using the Self-consistent Ornstein-Zernike approximation (SCOZA)
one assumes that the direct correlation function is the same as for
long-range forces outside hard cores, i.e., $-\beta{\tilde\psi}(k)$,
and that there is an additional term that contains a free parameter
determined from the consistency requirements of SCOZA.  Traditionally
this has been used to replace the temperature with an effective one,
but here we will rely on the parameter $z$ already introduced above.
The SCOZA equation for $D$-dimensional spins ($D\rightarrow\infty$)
connects transverse correlations and internal energy, i.e., with the
substitution $u=m^2$
\begin{equation}
\frac{\partial}{\partial\beta}\left(\frac{\beta}{\chi_\perp}\right)=\frac{\partial}{\partial\beta}\left(\frac{\beta H}{m}\right)=\frac{1}{m}\frac{\partial U}{\partial m}=2\frac{\partial U}{\partial u}.
\label{18}
\end{equation}
Inserting from Eqs.~(\ref{8}) and (\ref{9}) the SCOZA equation becomes
\begin{equation}
\nonumber
\frac{\partial}{\partial\beta}\left(\frac{\beta}{z}-\beta\right)=-1-\frac{1}{\beta}\frac{\partial}{\partial u}\left(P(z)-1\right)
\end{equation}
or ($P'(z)=dP/dz$)
\begin{equation}
\beta\frac{\partial z}{\partial\beta}-\frac{z^2}{\beta}P'(z)\frac{\partial z}{\partial u}=z
\label{19}
\end{equation}
whose equations for the characteristics are
\begin{equation}
\frac{d\beta}{\beta}=-\frac{\beta\,du}{z^2P'(z)}=\frac{dz}{z}.
\label{20}
\end{equation}
One solution of these equations is
\begin{equation}
\mu_e=\frac{z}{\beta}=C_1.
\label{21}
\end{equation}
The other solution follows from
\begin{equation}
\nonumber
du=-\frac{z}{\beta}P'(z)\,dz=-C_1P'(z)\,dz
\end{equation}
as
\begin{equation}
C_1P(z)=C_2-u.
\label{22}
\end{equation}
Comparing with the solution of the generalized mean-spherical model
(GMSM) above one finds that Eqs.~(\ref{21}) and (\ref{22}) are
identical to the exact solution (\ref{6}) and (\ref{14}) with
constants of integration
\begin{equation}
C_1=f(n)\quad {\rm and}\quad C_2=n.
\label{23}
\end{equation}
Thus for a given reference system the solution of the SCOZA problem
will reproduce the exact result (\ref{17}).

\section{HRT for the model.}
\label{sec5}

For the HRT we will also use expression (\ref{10}) for the correlation
function where $z$ again is the free parameter. By adding interaction
at wavevector $k=Q$ for decreasing $Q$ while keeping $\beta$ constant,
one obtains the equation \cite{reinerhoye05}
\begin{equation}
\frac{\partial I}{\partial Q}=4\pi CQ^2\ln{[1-z{\tilde\psi}(Q)]}\quad{\rm with}\quad C=\frac{1}{2}\frac{1}{(2\pi)^3}
\label{24}
\end{equation}
for space dimensionality $d=3$. The $I=-\beta f=L-\beta Hm$ where $f$
is Helmholtz free energy per spin while $L=-\beta g$ where $g$, which
also appears on the left hand side of Eq.~(\ref{2}), is the Gibbs free
energy per spin. For the transverse susceptibility we then have
($u=m^2$)
\begin{equation}
\frac{\beta}{z}-\beta=-\frac{1}{m}\frac{\partial I}{\partial m}=-2\frac{\partial I}{\partial u}.
\label{25}
\end{equation}
With (\ref{24}) inserted we get the HRT self-consistency equation
\begin{eqnarray}
\nonumber
\frac{\partial}{\partial Q}\left(\frac{\beta}{z}-\beta\right)=-2\frac{\partial}{\partial u}\left(\frac{\partial I}{\partial Q}\right)\\
\frac{\partial}{\partial Q}\left(\frac{\beta}{z}\right)=-\frac{\beta}{z^2}\frac{\partial z}{\partial Q}=4\pi\cdot2CQ^2\frac{{\tilde\psi}(Q)}{1-z{\tilde\psi}(Q)}\frac{\partial z}{\partial u}.
\label{26}
\end{eqnarray}
Its equations for the characteristics are
\begin{equation}
\frac{z}{\beta}\,dQ=\left[4\pi\cdot 2CQ^2\left(\frac{1}{1-z{\tilde\psi}(Q)}-1\right)\right]^{-1}\,du=\frac{dz}{0}.
\label{27}
\end{equation}
Again the solution is
\begin{eqnarray}
\nonumber
\mu_e=\frac{z}{\beta}=C_1\\
C_1 P(z)=C_2-u
\nonumber
\end{eqnarray}
where now in the integrand in integral (\ref{5}) for $P(z)=P(z,Q)$ the
$\tilde{\psi}(k)$ is replaced by 0 for $0<k<Q$. Comparing one sees
that this is nothing but Eqs.~(\ref{21}) and (\ref{22}).

\section{Alternative method.}
\label{sec6}

By requiring consistency between Helmholtz free energy and internal
energy when changing both $\beta$ and $Q$ Reiner and H{\o}ye obtained
a more general solution beyond the one of the mean spherical
approximation (MSA). They found \cite{reinerhoye05} 
\begin{eqnarray}
\nonumber
I&=&-C\int\ln(1-\mu_e \beta{\tilde\psi}(k))\,d{\bf k}-\sum\limits_{n=1}^\infty \frac{n}{n+1}A_n K^{n+1}\\
\mu_e&=&\mu+\sum\limits_{n=1}^\infty A_n K^n 
\label{28}
\end{eqnarray}
where $K=J/\mu_e$ and $P(z)=1+2J$. The coefficients $A_n$ do not
depend upon $\beta$ and $Q$. Note that here the $I$ does not contain
the reference system and mean field terms.  This expression will also
hold in the present case when imposing consistency with the
compressibility, but now the $A_n$ will depend upon the boundary
condition at $\beta=0$. We find by use of (\ref{28})
\begin{equation}
\beta H=\beta H_0(m)-\frac{\partial I}{\partial m}=\beta H_0(m)-\sum\limits_{n=1}^\infty \frac{1}{n+1}\frac{\partial A_n}{\partial m} K^{n+1}
\label{29}
\end{equation}
where $H_0$ is the reference system plus the mean field
contributions. So for the transverse susceptibility (\ref{8}) we get
($\mu_e=z/\beta$)
\begin{equation}
\frac{\beta H}{m}=\frac{1}{\mu_e}-\beta=\frac{1}{\mu}-\beta-2\left[\frac{\partial\mu}{\partial u}K+\sum\limits_{n=1}^\infty \frac{1}{n+1}\frac{\partial A_n}{\partial u} K^{n+1}\right].
\label{30}
\end{equation}
This equation together with (\ref{28}) determines $\mu_e$, i.~e., the
coefficients $A_n$ for given $\mu$ can be found by iteration by
comparing equal powers of $K$. However, this problem can be
transformed into the solution of a differential equation.  So by
choosing $u$ and $K$ as independent variables Eqs.~(\ref{28}) and
(\ref{30}) can be differentiated with respect to $u$ and $K$
respectively to obtain
\begin{equation}
\frac{1}{\mu_e^2}\frac{\partial \mu_e}{\partial K}=2\frac{\partial \mu_e}{\partial u}
\label{31}
\end{equation}
This partial differential equation has the solution
\begin{equation}
\mu_e=C_1 \quad{\rm and} \quad 2C_1^2 K=(C_2-C_1)-u.
\label{32}
\end{equation}
With $C_1(1+2C_1 K)=C_1 P(z)$ this is again solution (\ref{21}) and
(\ref{22}).

\section{Combined SCOZA and HRT.}
\label{sec7}

In Ref.~\onlinecite{reinerhoye05} consistency between free energy and
internal energy was used to determine a single free parameter.  This
gave rise to first order partial differential equations whose
properties were studied more closely.  Now we will require
thermodynamic consistency with the compressibility route, too, so that
a second free parameter can be determined.

Thus, to be general, consider a function $\Psi(\beta, Q,m)$ which is
determined via two free but unknown parameters
\begin{eqnarray}
z&=&z(\beta, Q, m),
\nonumber\\
\nu&=&\nu(\beta, Q, m).
\label{33}
\end{eqnarray}
For the determination of $\Psi$, $z$ and $\nu$, the derivatives of
$\Psi$ are given by known functions of $\beta$, $Q$, $m$, $z$, and
$\nu$ as
\begin{eqnarray}
\Psi_\beta&=&X=X(\beta,Q,m,z,\nu),
\nonumber\\
\Psi_Q&=&Y=Y(\beta,Q,m,z,\nu),
\label{34}\\
\Psi''&=&Z=Z(\beta,Q,m,z,\nu).
\nonumber
\end{eqnarray}
Here and below the subscripts mean partial derivatives with respect to
$\beta$ and $Q$ etc.~while the double prime means second derivative
with respect to magnetization $m$. For the GMSM the latter is replaced
by the first derivative with respect to $u=m^2$.

By differentiation with respect to $\beta$ and $Q$ we now get
\begin{eqnarray}
d\Psi_\beta&=&X_\beta\,d\beta+X_Q\,dQ+X_m\,dm+X_z\,dz+X_{\nu}\,d\nu,
\nonumber\\
d\Psi_Q&=&Y_\beta\,d\beta+Y_Q\,dQ+Y_m\,dm+Y_z\,dz+Y_{\nu}\,d\nu,
\label{35}\\
d\Psi''&=&Z_\beta\,d\beta+Z_Q\,dQ+Z_m\,dm+Z_z\,dz+Z_{\nu}\,d\nu,
\nonumber
\end{eqnarray}
where subscripts indicate partial derivatives with respect to $z$ and
$\nu$. With three unknowns $\Psi$, $z$ and $\nu$ the set of equations
(\ref{34}) represents a rather complex problem. We then note as in
Ref.~\onlinecite{reinerhoye05} that use of the identity $\partial
\Psi_\beta/\partial Q =\partial\Psi_Q/ \partial\beta$ will simplify
this, and we first obtain
\begin{equation}
X_Q+X_z z_Q+X_{\nu} \nu_{Q}=Y_\beta+Y_z z_\beta+Y_{\nu} \nu_{\beta}.
\label{36}
\end{equation}
Further
\begin{eqnarray}
\frac{\partial\Psi''}{\partial\beta}&=&Z_\beta+Z_z z_\beta+Z_{\nu} \nu_{\beta}=X'',
\nonumber\\
\frac{\partial\Psi''}{\partial Q}&=&Z_Q+Z_z z_Q+Z_{\nu} \nu_{Q}=Y''
\label{37}
\end{eqnarray}
or
\begin{eqnarray}
\nu_{\beta}&=&\frac{1}{Z_{\nu}}\left(X''-Z_z z_\beta-Z_\beta\right),
\nonumber\\
\nu_{Q}&=&\frac{1}{Z_{\nu}}\left(Y''-Z_z z_Q-Z_Q\right).
\label{38}
\end{eqnarray}
These expressions can be used to substitute the $\nu_\beta$ and
$\nu_Q$ in (\ref{36}), and by some rearrangement the following
equation is obtained
\begin{eqnarray}
&(Z_{\nu} X_z-X_{\nu}Z_z)z_Q-(Z_{\nu} Y_z-Y_{\nu}Z_z)z_\beta+&\nonumber\\
&X_{\nu}Y''-Y_{\nu}X''+Z_{\nu}(X_Q-Y_\beta)-Z_Q X_{\nu}+Z_\beta Y_{\nu}=0.
\label{39}
\end{eqnarray} 

The previous one-parameter approximations can be recognized in
Eq.~(\ref{36}) when $\nu$ is considered constant. Then with $z$ as the
free parameter and $\Psi=I=-\beta f$ where $f$ is free energy per
particle, one finds from Eq.~(\ref{38}) that $\nu_{\beta}=0$ is the
SCOZA equation, $\nu_{Q}=0$ is the HRT equation, while the remaining
(non-zero) terms of (\ref{36}) give the consistency between the free
energy and internal energy routes considered in
Ref.~\onlinecite{reinerhoye05}.  But in general these various
consistencies give different $z$ so (\ref{36}) itself will not be
solved by using one of these, except for the GMSM considered in this
work. Since we know the exact solution for the GMSM it should be
possible to revover it directly from (\ref{36}) and (\ref{39}) using a
pair correlation function containing two free parameters.

To obtain the resulting HRT-SCOZA equation the $X''$ and $Y''$ must be
evaluated. In the usual case we then have
\begin{eqnarray}
X'&=&X_m+X_z z_m+X_{\nu}\nu_{m},
\nonumber\\
X''&=&X_{mm}+2X_{mz}z_m+2X_{m\nu}\nu_{m}+X_{zz}z_{m}^2 +
\nonumber\\
&&2X_{z\nu}z_m \nu_{m}+X_{\nu \nu}\nu_{m}^2+X_z z_{mm}+X_{\nu}\nu_{mm}
\label{40}
\end{eqnarray}
with similar expression for $Y''$ with $X$ replaced by $Y$. One notes
that the $\nu_{mm}$ term will cancel when this is inserted in
(\ref{39}).  Thus the resulting HRT-SCOZA equation becomes a
second-order partial differential equation for $z$ with coefficients
that depend on $\nu$ and its first-order derivatives.  This can then
be treated iteratively, by starting with some approximate $\nu$,
solving for $z$, and updating $\nu$ according to Eq.~(\ref{38}).  Note
that $\nu_{m}=0$ for $m=1/2$ due to the symmetry of lattice gases if
we identify $\nu$ with $\mu_e$ as we will do below.  Its influence may
therefore be only perturbing in the updating process and thus not
crucial for the problem of performing a numerical solution.  Anyway,
here we will not try to pursue this question or try to analyse the
properties of the general HRT-SCOZA equation any further. Instead we
focus on the simplified situation with the GMSM to show how the
HRT-SCOZA equation solves this problem. As mentioned earlier the
susceptibility is then replaced by the transverse susceptibility. As
in Secs.~\ref{sec4} and \ref{sec5} we then put $u=m^2$ to get
\begin{eqnarray}
X''&\rightarrow2\frac{\partial X}{\partial u}=2[X_z z_u+X_{\nu}\nu_{u}+X_u]
\nonumber\\
Y''&\rightarrow2\frac{\partial Y}{\partial u}=2[Y_z z_u+Y_{\nu}\nu_{u}+Y_u].
\label{41}
\end{eqnarray}
For (\ref{39}) this amounts to the substitution
\begin{equation}
X_{\nu}Y''-Y_{\nu}X''\rightarrow2[(X_{\nu}Y_z-Y_{\nu}X_z)z_u+X_{\nu}Y_u-Y_{\nu}X_u],
\label{42}
\end{equation}
where now the $\nu_{u}$ term cancels. Thus we are left with a first
order partial differential equation for $z$ with free variables
$\beta$, $Q$ and $u$. But $\nu$, that is determined via (\ref{38}), is
still present in the coefficients of (\ref{39}).

\section{Two-parameter pair correlation function.}
\label{sec8}

A simple way to introduce two parameters in the correlation function is to modify Eq.~(\ref{10}) into
\begin{equation}  
{\tilde\Gamma_\perp}(k)
=\frac{\tilde{\Sigma}(k)}{1-\tilde{\Sigma}(k)\beta\tilde{\psi}(k)}
=\frac{\nu}{1-z\tilde{\psi}(k)}
\label{43}
\end{equation}
for $k\geq Q$ and $k=0$ (i.e., ${\tilde\Gamma_\perp}(k)=\tilde{\Sigma}(k)$
for $0<k<Q$). This means that the ``self-energy'' function is (for all
$k$)
\begin{equation}  
\tilde{\Sigma}(k)=\frac{\nu}{1-(z-\nu \beta)\tilde{\psi}(k)}.
\label{44}
\end{equation}
This assumed form of the correlation function for continuum spins with
two free parameters $\nu$ and $z$ can also be used for continuum
fluids and their lattice gas version too. Thus various HRT-SCOZA
expressions we derive in this section are also valid in the latter
cases before we again specialize to the GMSA below Eq.~(\ref{57}).

An interesting feature of expression (\ref{43}) is the adjustable
amplitude $\nu$ to which the internal energy is proportional. This
may influence critical properties. For SCOZA there is a generalized
kind of scaling \cite{hoyepinistell00}. The independence of $\nu$ from
$z$ may change this. Note that here the $\nu$ is not tied to a core
condition which we here omit for simplicity. Such an omission may not
be crucial for qualitative properties. Anyway, at least for SCOZA
itself, the core condition is not crucial in this respect
\cite{borgehoye98}.

With ${\tilde\Gamma_\perp}(k)$ given above we can now evaluate the
quantities that enter the HRT-SCOZA equation. With $\Psi=I=- \beta f$
where $f$ is Helmholtz free energy per particle we have
\cite{reinerhoye05}
\begin{eqnarray}
X&=&\frac{\partial I}{\partial\beta}=\frac{\nu}{z}J(z)+\frac{1}{2}m^2
\label{46}\\
Y&=&\frac{\partial I}{\partial Q}=4\pi CQ^2\ln{(1-\tilde{\Sigma}(Q)\beta\tilde{\psi}(Q))}
\nonumber\\
&=&4\pi CQ^2[\ln{(1-z\tilde{\psi}(Q))}-\ln{(1-(z-\nu \beta)\tilde{\psi}(Q))]}
\label{47}\\
Z&=&2\frac{\partial I}{\partial u}=-\frac{1}{{\tilde\Gamma_\perp}(0)}=-\frac{1-z}{\nu}\qquad (u=m^2)
\label{48}
\end{eqnarray}
with, for given $Q$,
\begin{equation}  
J(z)=\frac{1}{2}(P(z)-1)=C\int\limits_{k>Q} \frac{z\tilde{\psi}(k)}{1-z\tilde{\psi}(k)}\,d{\bf k}.
\nonumber
\end{equation}

Here $X=-U$ is a modification of expression (\ref{9}) for the internal
energy $U$, the $Y$ is a modification of expression (\ref{24}), while
$Z$ is the corresponding modification of expression (\ref{25}). From
this we obtain the partial derivatives
\begin{eqnarray}
Y_\beta&=&-\nu L(Q,\Delta z),
\nonumber\\
Y_z&=&-L(Q,z)+L(Q,\Delta z),
\label{49}\\
Y_{\nu}&=&-\beta L(Q, \Delta z), \qquad Y_u=0
\nonumber
\end{eqnarray}
where $\Delta z=z-\nu \beta$ and
\begin{equation}  
L(Q,z)=-\frac{1}{z}\frac{\partial J(z)}{\partial Q}=4\pi CQ^2\frac{\tilde{\psi}(Q)}{1-z\tilde{\psi}(Q)}.
\label{50}
\end{equation}
Further, with $J'(z)=-\partial J(z)/\partial z$
\begin{eqnarray}
X_Q&=&-\nu L(Q, z),
\nonumber\\
X_z&=&\nu \frac{\partial}{\partial z}\left(\frac{J(z)}{z}\right)=-\frac{\nu}{z^2}J(z)+\frac{\nu}{z}J'(z),
\label{51}\\
X_{\nu}&=&\frac{1}{z} J(z),\qquad X_u=\frac{1}{2},
\nonumber
\end{eqnarray}
and finally
\begin{eqnarray}
Z_\beta&=&0, \qquad Z_Q=0,
\nonumber\\
Z_z&=&\frac{1}{\nu}, \qquad\,\, Z_{\nu}=\frac{1-z}{\nu^2}.
\label{52}
\end{eqnarray}
For the GMSM case where (\ref{42}) is used for the $X''$ and $Y''$
terms we now can evaluate the coefficients of the HRT-SCOZA equation
(\ref{39}) to obtain
\begin{equation}  
Az_Q-Bz_\beta+2Cz_u+D=0
\label{53}
\end{equation}
where the coefficients are
\begin{eqnarray}
A&=&Z_{\nu}X_z-X_{\nu}Z_z=A_1+A_2,
\nonumber\\
A_1&=&-\frac{1}{\nu z^2}J(z),\qquad A_2=\frac{1-z}{\nu z}J'(z),
\label{54}
\end{eqnarray}
\begin{eqnarray}
B&=&Z_{\nu}Y_z-Y_{\nu}Z_z=B_1+B_2,
\nonumber\\
B_1&=&\frac{1-z}{\nu^2}L(Q,z),\qquad B_2=\frac{1-\Delta z}{\nu^2}L(Q,\Delta z).
\label{55}
\end{eqnarray}
With (\ref{42}) we have
\begin{eqnarray}
C&=&X_{\nu}Y_z-Y_{\nu}X_z=C_1+C_2,
\nonumber\\
C_1&=&\frac{1}{z}J(z)\left[-L(Q,z)+\frac{\Delta z}{z^2}L(Q,\Delta z)\right],
\nonumber\\ 
C_2&=&\frac{\beta\nu}{z}J'(z)L(Q,\Delta z).
\label{56}
\end{eqnarray}
Finally,
\begin{eqnarray}
D&=&Z_{\nu}(X_Q-Y_\beta)-Z_Q X_{\nu}+Z_\beta Y_{\nu}+2(X_{\nu} Y_u-Y_{\nu} X_u)
\nonumber\\
&=&Z_{\nu}(X_Q-Y_\beta)-Y_{\nu} =D_1+D_2,
\nonumber\\
D_1&=&-\frac{1-z}{\nu}L(Q,z)=\nu B_1,
\nonumber\\ 
D_2&=&\frac{1-\Delta z}{\nu}L(Q,\Delta z)=\nu B_2.
\label{57}
\end{eqnarray}

Now in the GMSM the the solution to be expected yields $\nu=z/\beta$.
This suggests to replace $z$ with a new variable $\mu=z/\beta$, which
simplifies the remaining analysis since then the $D$-terms will join
$B$-terms, and Eq.~(\ref{53}) becomes
\begin{equation}  
E_1+E_2+E_3=0
\label{58}
\end{equation}
with
\begin{eqnarray}
E_1&=&A_2 \mu_Q-B_1 \mu_\beta=\beta\frac{1-z}{z^2}[J'(z)\mu_Q+\beta L(Q,z)\mu_\beta],
\nonumber\\
E_2&=&A_1\mu_Q+2C_1\mu_u=-\frac{1}{z}J(z)\left[\frac{\beta}{z^2}\mu_Q+2L(Q,z)\mu_u\right],
\nonumber\\
E_3&=&-B_2\mu_\beta+2C_2\mu_u=-\frac{1}{z^2}L(Q,\Delta z)[-\beta^2 \mu_\beta+2z^2 J'(z)\mu_u].
\label{59}
\end{eqnarray}

Now one notes that $E_3=0$ is the SCOZA equation (\ref{19}) as
$P(z)=1+2J$ and $z=\beta\mu$. Likewise $E_2=0$ is the HRT equation
(\ref{26}). These equations have both the common GMSM solution given
by (\ref{21}) and (\ref{22}). Noting further that
\begin{equation}
\frac{1}{1-z}E_1=-\frac{z J'(z)}{J(z)}E_2+\frac{L(Q,z)}{L(Q,\Delta z)}E_3, 
\nonumber
\end{equation}
it follows that the GMSM solution also solves $E_1=0$, and by that it
solves the HRT-SCOZA equation (\ref{58}) too. Here it can be noted
that $E_1=0$ is nothing but consistency between the internal energy
and free energy routes investigated in Ref.~\onlinecite{reinerhoye05}.
One should expect that Eq.~(\ref{58}) have other solutions too with a
third constant of integration $C_3$ besides the two in
Eq.~(\ref{22}). But in any case the GMSM solution is sufficient here
as it can be adjusted into the reference system boundary conditions
(\ref{12}) or (\ref{23}).

Finally we have to show that $z=\beta \nu$ fulfills Eq.~(\ref{38})
too. With $\mu=\nu=z/\beta$ and Eqs.~(\ref{41}) and
(\ref{49})--(\ref{52}) we get
\begin{eqnarray}
\mu_\beta&=&\frac{1}{Z_{\nu}}[2(X_z z_u+X_{\nu} \mu_u+X_u)-Z_z z_\beta -Z_\beta]=\frac{1}{1-z}[2\mu^2J'(z)\mu_u-z\mu_\beta)]
\nonumber\\
\mu_\beta&=&2\mu^2 J'(z)\mu_u
\label{60}
\end{eqnarray}
and
\begin{eqnarray}
\mu_Q&=&\frac{1}{Z_{\nu}}[2(Y_z z_u+Y_{\nu} \mu_u+Y_u)-Z_z z_Q -Z_Q]=\frac{\mu z}{1-z}[-2L(Q,z)\mu_u-\frac{1}{\mu}\mu_Q)]
\nonumber\\
\mu_Q&=&-2\mu zL(Q,z)\mu_u.
\label{61}
\end{eqnarray}
And Eqs.~(\ref{60}) and (\ref{61}) are nothing but the SCOZA and HRT
equations $E_3=0$ and $E_2=0$ as given by (\ref{59}). Thus altogether
we have shown in detail how the GMSM solves the unified HRT-SCOZA
equations. We have reason to believe that this demonstration of a
model that can be solved exactly will be useful for the possible
solution of the HRT-SCOZA problem more generally where the second
derivatives of Eq.~(\ref{40}) should be used. Also other assumptions
for the correlation function different from the simple expression
(\ref{43}) may then be useful or may be needed.

\section{Conclusions}

In the present work general equations for the unified HRT-SCOZA
problem have been established using a simple form of the pair
correlation function containing two free parameters.  To analyse the
problem in more detail we have considered an exactly solvable model,
the MSM and its extension the GMSM that we introduce.  This
generalization is also natural in so far as the GMSM is merely a more
general solution of the same HRT-SCOZA equations.  The reference
system boundary conditions determines the resulting solution.  The
SCOZA and HRT problems for the GMSM are first considered separately,
and then they are combined.  By analysis of the unified HRT-SCOZA it
is shown how it can reproduce the known exact solution of the GMSM:
Given correct boundary conditions and a suitable parameterization of
the correlation function, HRT-SCOZA successfully traces the evolution
of the free parameters.  We expect the analysis of how the HRT-SCOZA
works for this special case to be useful for the more general
situation of possibly solving the unified HRT-SCOZA problem.

\acknowledgments

AR gratefully acknowledges financial support from \textit{Fonds zur
F\"orderung der wissenschaftlichen Forschung (FWF)} under
project~J2380-N08.

\end{document}